\documentclass[12pt,a4wide]{article}
\usepackage{a4wide}
\usepackage{textcomp}
\usepackage{hyperref}
\usepackage{amstext,amsmath,amssymb,mathrsfs}
\usepackage{float}
\usepackage{cite}
\newif\ifshowcitations\showcitationsfalse%
\newif\ifshowlinks\showlinksfalse%
\usepackage[normalem]{ulem}
\usepackage{color}
\usepackage{mathtools}
\usepackage{enumitem}
\usepackage[utf8]{inputenc}
\newcommand{\lm}[1]{\textcolor{red}{[LM: #1]}}
\newcommand{\md}[1]{\textcolor{blue}{[MRD: #1]}}
\newcommand\setItemnumber[1]{\setcounter{enumi}{\numexpr#1-1\relax}}
\ifshowlinks%
  \usepackage[
         colorlinks=true,
         urlcolor=blue,       
         ]{hyperref}
  \newcommand*{\inspireurl}[1]{\\\href{#1}{INSPIRE-HEP entry}}
\else
  \makeatletter
  \newcommand*{\inspireurl}[1]{\@bsphack\@esphack}
  \makeatother
\fi
\ifshowcitations%
  \newcommand*{\citations}[1]{\\* #1}
\else
  \makeatletter
  \newcommand*{\citations}[1]{\@bsphack\@esphack}
  \makeatother
\fi

\def\K{K} 
\def\dimK{k} 
\def\Mod{\mbox{Mod}} 
\def\Isom{\mbox{Isom}} 
\def\B{{B}} 
\def\Veff{{V_{\text{eff}}}} 
\def\Seff{\mathcal{S}} 
\def\m{\phi} 
\def\boldclass{\bf\sf}
\def\theory{{\boldclass T}}
\def\Mtheory{{\boldclass M}}
\def\Ftheory{{\boldclass F}}
\def\IIatheory{{\boldclass IIa}}
\def\IIbtheory{{\boldclass IIb}}
\def\Itheory{{\boldclass I}}
\def\HEtheory{{\boldclass HE}}
\def\HOtheory{{\boldclass HO}}
\def\GR{{\boldclass GR}}
\def\O7{{\boldclass O7}}
\def\YM{{\boldclass YM}}
\def\ED3{{\boldclass ED3}}
\def\D3{{\boldclass D3}}
\def\MD7{{\boldclass D7}}
\def\pq7{{\boldclass (p,q)7}}
\def\Ed(-1){{\boldclass ED(-1)}}
\def\Sigmatheory{\Seff} 
\def\CalabiYau{{CY}}
\def\Etheory{{\boldclass E}}
\def\EFT{{\boldclass EFT}}
\def\Vac{\text{Vac}}
\newcommand{\term}{}

\def\Vol{\mbox{Vol}}
\def\BC{\mathbb{C}}

\def\BZ{\mathbb{Z}}
\def\BR{\mathbb{R}}

\def\CM {{\cal M}}

\newcommand{\eq}[1]{Eq.~(\ref{eq:#1})}

\newcommand{\Tr}{{\rm Tr}}  

\def\one{{\hbox{ 1\kern-.8mm l}}}

\def\vol{{\rm vol\,}}

\begin{document}
\pagestyle{plain}
\begin{center}
\medskip
{\bf\LARGE{Compactification of Superstring Theory}}
\vspace{10mm}
\end{center}

\begin{center}
Michael R.~Douglas$^{a}$ and Liam McAllister$^{b}$
\end{center}

\begin{center}
\vspace{0.15 cm}
{\fontsize{11}{30}
\noindent\textsl{$^{a}$CMSA, Harvard University, Cambridge, MA 02138, USA}}\\
\vspace{0.15 cm}
{\fontsize{11}{30}
\noindent\textsl{$^{b}$Department of Physics, Cornell University, Ithaca, NY 14853, USA}}\\
\end{center}

\vspace{1cm}
\noindent
We give a mathematical perspective on string compactifications.
Submitted as a chapter in the Encyclopedia of Mathematical Physics.

\vspace{8.25cm}
\noindent\today
\newpage

\tableofcontents
\newpage

\section{Introduction}

Superstring theories and M-theory, at present the best candidates for
quantum theories unifying gravity, matter, and Yang-Mills fields,
are directly formulated in ten and eleven space-time dimensions.  To
obtain a candidate theory of our four-dimensional universe, one must
find a solution of one of these theories whose low-energy physics
contains the well-established Standard Model of particle physics coupled to Einstein's general relativity.

The standard paradigm for finding such solutions is
compactification, along the lines originally proposed by Kaluza and
Klein in the context of higher-dimensional general relativity.
One postulates that the underlying $D$-dimensional space-time is
a product of four-dimensional Minkowski space-time with a
$(D-4)$-dimensional compact and small Riemannian manifold $\K$.
One then finds that low-energy physics effectively
averages over $\K$, leading to a
four-dimensional effective field theory (EFT) whose field content and Lagrangian are determined
by
the topology and geometry of $\K$.

A central problem in string compactification is:
\begin{center}
\emph{Find all compactifications $\mathscr{C}$
and the corresponding effective field theories
$\EFT(\mathscr{C})$}\,.
\end{center}

The subsequent sections are as follows.
In \S\ref{s:gen} we discuss the general problem of Kaluza-Klein compactification,\footnote{In this article we largely restrict attention to
compactifications based on compact spaces $\K$ that are Ricci flat.  There is an
equally important class in which $\K$ has positive curvature: these
lead to anti-de Sitter space-times and are discussed in the articles
on AdS/CFT.} we explain the data $\mathscr{C}$ characterizing a compactification, and we give a number of examples.
In \S\ref{sec:eft} we describe how effective field theories are obtained from
compactifications.
In \S\ref{s:susy} we consider supersymmetric compactifications of string theory and M-theory.
In \S\ref{s:typeII} we discuss type IIB orientifolds with flux in some detail.
We conclude by indicating some open problems.

\section{Kaluza-Klein compactification}
\label{s:gen}

In Kaluza-Klein (KK) compactification, we are given a $D$-dimensional field theory
$\theory$, usually general relativity (GR) coupled to matter
fields\footnote{We refer to all fields other than the gravitational and gauge fields as ``matter'', regardless of their spin.}
which we schematically denote $\varphi$.
We choose a $\dimK$-dimensional manifold $\K$ with Riemannian metric $G$,
called the compactification or internal manifold.

A compactification of $\theory$ on $\K$ consists of
a $D$-dimensional space-time $\CM_D$ that is a fibration over a $d$-dimensional space-time $\CM_d$ ($d=D-k$),
\begin{eqnarray}
\K\rightarrow \CM_D\rightarrow \CM_d\,,
\end{eqnarray}
and whose metric is a warped product,
\begin{equation} \label{eq:ansatz}
ds^2 = w^{4/d}\,g_{\mu\nu} dx^\mu dx^\nu + G_{ij} dy^i dy^j + A_{\mu i} dx^\mu dy^i\,,
\end{equation}
together with a matter field configuration that satisfies
the equations of motion of $\theory$ and any topological constraints.
In \eqref{eq:ansatz}, $g_{\mu\nu}$ is the metric on $\CM_d$, $G_{ij}$ is the Riemannian metric on
$\K$, $A_{\mu i}$ are mixed metric components, and $w$ is a positive real-valued function on $\K$ called the warp factor.

Compactification can also be defined for $\theory$ a string theory or M-theory.
In the limit of low energies, small curvatures, and weak coupling, the
various string theories and M-theory reduce to the ten-dimensional and eleven-dimensional maximal supergravity theories, supplemented with a few essential corrections from string theory.
Thus, one can study
string/M-theory compactification
by starting with a compactification of
a related supergravity theory, and then applying modifications that we will discuss.\footnote{Intrinsically string-theoretic approaches to compactification are possible in some cases, but in this article we will focus on compactifications based on supergravity,
leaving many stringy topics to the articles on conformal field theory, topological string
theory, and so on.}
One such modification is to include ``branes,'' which we denote $\B$.
As we explain in the examples below, branes can be understood as lower-dimensional theories that introduce source terms in the equations of motion of $\theory$, and that have fields and equations of motion of their own.

A maximally symmetric compactification, sometimes called a vacuum, is one for which $\CM_d$ is a
solution of the vacuum $d$-dimensional Einstein equations with maximal isometry group.  The choices are
Minkowski space,
de Sitter space (dS),
and anti-de Sitter space (AdS),
in which the $d$-dimensional cosmological constant $\Lambda$ is zero, positive, and negative, respectively.
To respect the space-time isometries,
we require $A_{\mu i}=0$ and take the configuration (expectation values) of the fields $\varphi$ to be invariant up to gauge transformation.

To summarize, the data of a compactification is
$\mathscr{C}=(\theory,K,\B,G,\varphi,\Lambda,w)$.  Given this data, we could in principle compute the outcome of
any $d$-dimensional experiment performed in the universe it models.  Conversely, given experimental data
one would like to constrain the compactification.  This is usually done using the EFT formalism discussed in \S\ref{sec:eft}.

\subsection*{Examples }
\label{ss:examples}

We begin with basic examples illustrating the possibilities in compactification of field theories and string theories.
\begin{enumerate}
    \item Let $\theory=\GR_D$, pure $D$-dimensional general relativity with
zero cosmological constant, so that Einstein's equations reduce to the condition of Ricci flatness.
Assuming \eq{ansatz}, one can show that the metric $G$ on $\K$ must be Ricci flat and $w$ must be constant,
leading to a Minkowski solution.
The simplest example is the torus $\K= T^\dimK$, while
examples of $\K$ without continuous isometry include the K3
manifold ($k=4$), the Calabi-Yau $n$-folds ($k=2n$), and manifolds of $G_2$ holonomy ($k=7$).

\item Let $\theory$ be Einstein-Maxwell theory and let $\K=S^2$. Then there are AdS compactifications with
$G$ the round metric and $\varphi$ a nonzero constant curvature connection.  The constant curvature $U(1)$ connections
on $S^2$ are classified by the first Chern class, so one has a family of solutions labeled by an element of $H^2(S^2,\mathbb{Z})=\mathbb{Z}$, the ``flux'', or equivalently by a
nonzero integer, the number of units of flux.

\item Supergravity theories (discussed in \S \ref{s:susy}) contain $(p+1)$-form field strengths,\footnote{
The convention $p+1$ is standard; the ``$+1$'' is related to the notation ``$3+1$'' for 3 space and 1 time dimensions.}
generalizing the Maxwell case where $p+1=2$.  These theories have AdS compactifications with
$\K=S^{p+1}$, again labeled by a non-zero integer (due to flux quantization).  There is a second
series of ``electric flux'' compactifications with $\K=S^{D-p-1}$, also labeled by a nonzero integer, which can be understood as having nonzero values of the Hodge dual $(D-p-1)$-form field strengths.

\item String theories and M-theory admit brane solutions of many types, generally in correspondence with the $(p+1)$-form and dual
$(D-p-1)\equiv (p'+1)$-form field strengths.  Geometrically the branes look like submanifolds of dimension $p+1$ and $p'+1$.
One can use a brane with $p+1\ge d$ in compactification by letting this submanifold be $\CM_d\times \Sigma$, where
 $\Sigma\subseteq\K$ is a $(p+1-d)$-dimensional submanifold.  This (and the analog with $p\rightarrow p'$)
is called ``wrapping a spacetime-filling brane on $\Sigma$.''  To satisfy the
equations of motion, the submanifold generally must be
volume-minimizing.\footnote{In cases where it is of interest to preserve supersymmetry, the submanifold generally must furthermore be calibrated, which implies volume-minimizing but not conversely.}
If there are multiple $(p+1)$-forms,
each brane is associated to a specific linear combination of them;
for example the Neveu-Schwarz-Neveu-Schwarz and Ramond-Ramond two-forms of $\theory=\IIbtheory$ can be
combined as $m B^{NS} + n B^{RR}$ and give rise to $(m,n)$ branes.
Branes often have additional matter as well, for example Dirichlet branes
in string theory carry gauge connections, which come with a choice of vector bundle topology.  The submanifolds $\Sigma$ along with these
topological choices are denoted $\B$. \label{ex:brane}

\item Let $\theory=\IIbtheory$ (type IIb superstring theory) with $D=9+1$ and take $d=3+1$.  This theory has Dirichlet 3-branes (D3-branes), which are objects
that embed into $(3+1)$-dimensional submanifolds of $D$-dimensional space-time.  Some compactifications of
$\IIbtheory$ include a number $N_{D3} \neq 0$
of D3-branes, each of
which embeds into $\CM_d \times \mbox{pt}$,
where $\mbox{pt}\in\K$ is a point.
In some cases one can begin with a compactification without D3-branes and change some of the discrete data, such as fluxes, to arrive at a
related but different compactification containing D3-branes.
We discuss consistency conditions for such changes in \S\ref{s:typeII}.
\end{enumerate}

In all of these examples, the compactifications come in families with discrete parameters, continuous parameters (called \emph{moduli}), or both.
In example 2, the first Chern class of the bundle on $S^2$ is a discrete parameter.
In example 5, the position of the D3-brane is a continuous parameter.
In both cases there are consistency conditions and equations of motion restricting the allowed parameter values.

In example 1, because the Ricci flatness condition is scale invariant, the overall scale of the metric is a continuous parameter.
More explicitly, given one Ricci flat metric $G_1$ we can construct a line of Ricci flat metrics
$\{G_\lambda \equiv \lambda^2 G_1 \,\forall \lambda\in(0,\infty)\}$.  Usually there are more moduli.
For example, Yau's theorem tells us that Ricci-flat K\"ahler metrics are parameterized by a choice of complex structure
and a choice of K\"ahler class.  The moduli spaces ${\mathcal M}$ of these metrics thus contain the factors
\begin{equation}\label{eq:mprod}
{\mathcal M} \supset {\mathcal M}_c \times {\mathcal M}_K\,,
\end{equation}
where ${\mathcal M}_c$ is the complex structure moduli space, with complex dimension given by the Hodge number $h^{k/2-1,1}$, and
${\mathcal M}_K$ is the K\"ahler moduli space, with real dimension given by $h^{1,1}$.
If $\theory$ has Yang-Mills fields
(as in the heterotic string), their solution space will typically have moduli.  If there are branes, these usually
come with moduli of the cycles they wrap and bundles they carry. These many choices typically lead to large families
of compactifications, all with the same topology of compactification manifold $\K$ and branes with bundles $\B$.

To discuss the next set of examples, we introduce the notation $\int_{\K,\B} \theory$ for the family of all compactifications with a fixed topology\footnote{The topology of $\K$ and $\B$ is fixed in the sense of ``stringy topology,'' which allows more continuous deformations than conventional topology.  We will give
an example in \S \ref{sec:iibsugra}.} of $\K$ and $\B$ (if $\B$ is omitted there are no branes).

\begin{enumerate}
\setItemnumber{6}

\item Compactification can be iterated.  If $\K\cong \K_1\times \K_2$ then one can first compactify on $\K_1$ to get a $(D-\dim K_1)$-dimensional
theory, which is then compactified on $\K_2$.  If both are small, the order does not matter, $\int_{\K_1}\int_{\K_2}\theory\cong\int_{\K_2}\int_{\K_1}\theory$.
There are also many results for fibrations $\K_1\rightarrow \K\rightarrow \K_2$ in which $\int_\K\theory \cong \int_{\K_2,\B} \theory'$ with
$\theory'=\int_{\K_1}\theory$ and the fibration structure and singularities determine the choice of $\B$.  As an example, take $\K_1=S^1$.
A smooth metric on $\K$ can have fibration singularities which look locally like
asymptotically locally flat metrics.  These are
represented in $\theory'$ as branes with $p+1=D-4$, and are usually called KK branes. \label{ex:fiber}

\item Symmetries of $\theory$ that act trivially on the compactification data lead to symmetries in the compactified theory.
For example, if $\theory$ includes $\GR$, the group of continuous isometries $\mbox{Isom}(K,G)$ will be a subgroup of the
gauge group of $\int_K\theory$.  If the compactification data is then varied in a way which breaks this symmetry, this can be
seen as spontaneous symmetry breaking.
In \S \ref{s:susy} we discuss the analog of this for supersymmetry.

\item All of these constructions generalize directly to $\K/\Gamma$ where $\Gamma$ is a freely acting finite isometry group.
If $\Gamma$ is not freely acting one may get what is called an orbifold compactification.  Whether this makes sense physically
depends on the resulting singularities.  As a rule of thumb, if the singularity can be obtained by taking a limit of noncompact
metrics of the same type (Ricci flat or special holonomy) as that on $\K$, then the orbifold will make sense.


\item String theories and M-theory admit special finite group actions which can be combined with $\Gamma$ in the quotient in example 8.
For example, the type II strings admit
$\BZ_2$ symmetries
$\Omega_{\text{ws}}$ that reverse the
orientation of the string worldsheet.  A quotient by the product $\Gamma^\prime$ of $\Omega_{\text{ws}}$
with a $\BZ_2$ isometry $\Gamma$ of $(\K,G)$ is called  an orientifold compactification.  Just as $K/\Gamma$ in example 8 can be thought
of as a generalized manifold $K'$, so too this quotient can be thought of as a generalized manifold.  The fixed loci of the action $\Gamma^\prime$ in $K$ are called orientifold planes, and are a special example of the branes introduced in example 4.

\item String theories and M-theory, and the associated supergravity theories supplemented by brane sources, admit solutions called flux compactifications that generalize example 3.  For example, with $\theory=\IIbtheory$, $K$ a Calabi-Yau $3$-fold, and taking a quotient $\Gamma^{\prime}$ for which the geometric involution $\Gamma$ is holomorphic, a compactification of $\theory$ on $K$ has AdS solutions containing quantized three-form field strengths, and in which the branes are D7-branes, O7-planes, and O3-planes.
While in the
flux compactifications on spheres of
of examples 2 and 3 the radii of curvature of AdS and $K$ are comparable, in more general flux compactifications these scales can be hierarchically different.  An example is presented in \S\ref{s:vacuumexample}.
\label{ex:IIbflux}

\end{enumerate}

\section{Effective field theory}\label{sec:eft}

Given a
compactification
$\mathscr{C}=(\theory,K,\B,G,\varphi,\Lambda,w)$
and an energy scale
$M\in \BR^+$ called the cutoff, the EFT of $\mathscr{C}$ with cutoff $M$, denoted here $\EFT(\mathscr{C},M)$,
is a $d$-dimensional field theory
that --- with a suitable choice of vacuum, as explained below --- reproduces all the observables of $\mathscr{C}$ at energy scales $E\le M$.
Thus, fixing the same $d$-dimensional
space-time metric $g_{\mu\nu}$, the
symmetries of $\mathscr{C}$ are the same as those of $\EFT(\mathscr{C},M)$, particles in the compactification with mass $m\le M$ all have associated fields in the EFT,
the $n$-point functions of these fields are the same, {\it etc.}

While the specifics of the EFT depend on the cutoff $M$,
for any prescribed cutoff $M' < M$ the renormalization group defines a map
\begin{equation}\label{eq:rgmap}
\text{RG}:\,\EFT(\mathscr{C},M) \to \EFT(\mathscr{C},M')\,,
\end{equation}
yielding an EFT with compatible properties.
One says that the degrees of freedom of $\EFT(\mathscr{C},M)$ with frequency $M'<\omega <M$ are `integrated out' via the map \eqref{eq:rgmap}.

\subsection{Effective action and effective potential}

To the extent that the EFT can be described by quantizing an action (for example, if $\hbar$ is small), there is a corresponding
effective action  $\Seff$ that furnishes a simple summary of the EFT.
To formulate $\Seff$ mathematically we specify a space
$\mathscr{T}$, known as the target space, with a Riemannian metric, and a gauge group $\mathscr{G}$ that acts by isometries on $\mathscr{T}$.
Maps
\begin{equation} \label{eq:Phidef}
    \Phi: \CM_d \to \mathscr{T}
\end{equation} are called field configurations, and we consider
an associated action functional $S:\{\Phi\} \to \BC$ that depends on the cutoff $M$.\footnote{
While $S$ is complex, physical consistency places additional requirements on the individual terms.  For example, the effective potential
defined below must be real.}
Then the data of an effective action $\Seff$ is $\Seff(\mathscr{G},\mathscr{T},S,M)$.\footnote{Fuller explanations are in the articles on quantum field theory and effective field theory.}

Constant field configurations $\Phi_{\text{const.}}$ are those that associate all of $\CM_d$ to one point in $\mathscr{T}$.  Restricting $S$ to constant field configurations and taking the cutoff $M \to 0$, we define
the \emph{effective potential} $\Veff$,
\begin{equation}
    \Veff:= S\bigl(\Phi_{\text{const.}},M=0\bigr): \mathscr{T} \to \BR\,.
\end{equation}
The effective potential is important because it determines the possible vacuum configurations, and so often it is useful to work with the data
\begin{equation}
\Seff(\mathscr{G},\mathscr{T},\Veff)\,.
\end{equation}

Introducing coordinates $\phi^i$ on $\mathscr{T}$,
a vacuum is a $\mathscr{G}$-equivalence class of critical points $\partial\Veff/\partial \phi^i=0$,
a metastable vacuum is a local minimum of $\Veff$, and a stable vacuum is a global minimum of $\Veff$.
The minima do not have to be strict, and thus there can be lines or other connected sets of vacua.
Denote the space of vacua of an EFT $\Etheory$ as
\begin{equation}
    \Vac(\Etheory) \subseteq \mathscr{T}\,.
\end{equation}

The {\it a priori} constraints on $\mathscr{G}$, $\mathscr{T}$ and $\Veff$ are rather weak,\footnote{
Supersymmetry leads to stronger constraints: see \S \ref{ss:superEFT}.
}
and to a first approximation can be freely specified: let $\Seff(\mathscr{G},\mathscr{T},\Veff)$ be the EFT
so defined.  Now, while not all EFTs can be obtained by quantizing an action, the Standard Model can
be, and there is a sense in which the generic EFT can be so obtained.
Thus, as stated in the introduction, one can summarize a large part of the problem of string compactification as: find all compactifications $\mathscr{C}$
and the corresponding EFTs,
\begin{equation}
\EFT(\mathscr{C})=\Seff(\mathscr{G},\mathscr{T},\Veff)\,.
\end{equation}

The vacua are so called because each one corresponds to a constant background field configuration that does not fluctuate
in the quantum theory, and so in general each vacuum leads to different physical predictions.\footnote{
This is actually only true in $d\ge 3$.}
We are now in a position to clarify the statements made at the start of \S\ref{sec:eft}:
given a compactification $\mathscr{C}$,
there is an $\Etheory=\EFT(\mathscr{C},M)$ and a choice of vacuum in $\Vac(\Etheory)$ that reproduces
the physics of $\mathscr{C}$.

\subsection{Moduli}\label{ss:moduli}

There is an important
correspondence
between the family or moduli space structure of a set of compactifications,
and $\mathscr{T}$ in the corresponding EFTs.  If there is a continuous deformation within the family
starting from $\mathscr{C}$, it must be realizable as a continuous motion within $\Vac\bigl(\EFT(\mathscr{C})\bigr)$.
In this sense, an EFT naturally corresponds to a family of compactifications, and this is the reason for
the notation $\int_{\K,\B} \theory$ introduced above.

Consider example 1 of \S \ref{ss:examples}.  The parameter $\lambda$ (the scale of the metric on $\K$)
translates into a scalar field $\phi_\lambda$ in $\int_\K \theory$.  All of the
dependence of observables on $\lambda$ factors through the expectation value of the field $\phi_\lambda$.
Likewise, if the solution space for the matter fields $\varphi$
has continuous parameters, or the brane configuration $B$ (such as the choice of $\Sigma$ in example \ref{ex:brane} of \S\ref{ss:examples}) has continuous
parameters, these will also translate into scalar fields.  In general
the equations of \theory\ couple these
choices.  In any case, the notation $\int_{K,B} \theory$ is meant to describe this single EFT,
which describes an entire connected set of compactifications.

In more complicated examples, the moduli space is the set of vacua $\Vac(\Etheory)$ embedded in a larger field space.
To illustrate, suppose there were precisely two Ricci flat metrics $G_1$, $G_2$ on the same $\K$.
Since they are connected in the space of all metrics, for sufficiently large cutoff $M$ they will correspond to
two points $\phi_1$, $\phi_2$ connected
by a path in field space, but now $\Veff$ will not be constant along this path.

The effective potential is a central aspect of example \ref{ex:IIbflux} of \S \ref{ss:examples}.  While getting a useful explicit expression
requires the methods discussed in \S\ref{s:vacuumexample}, the fact that fluxes contribute to the effective potential is quite simple to see.  Suppose
$\theory$ has a $(p+1)$-form field strength $F$
that enters the action through a term $\int_\K  F\wedge \star F$.
Then the equations of motion of $F$ are solved by harmonic forms $dF=d\star F=0$,
and such forms are uniquely determined by their class $[F] \in H^{p+1}(\K,\BR)$.
Suppose this class is nontrivial: physical consistency (Dirac quantization)
then requires that
$[F]\in H^{p+1}(\K,\BZ)$, so one has a a family of compactifications labeled by the discrete parameter $[F]$.
The energy of one of these, taking $F$ to be the harmonic form, is
\begin{equation} \label{eq:fluxe}
    E = \int_\K  F\wedge \star F\,,
\end{equation}
which depends on the metric $G$ on $\K$ through the Hodge star $\star$.  If there is a moduli space of metrics,
then one expects the energy to depend on the specific choice of metric.
For example, with $K$ a quotient of a Calabi-Yau threefold, as in example \ref{ex:IIbflux} of \S\ref{s:gen}, $G$ the Ricci-flat metric on $K$, $\theory=\IIbtheory$, and $p=2$, the energy \eqref{eq:fluxe} depends on the complex structure moduli of $G$.

\emph{Moduli stabilization} is the study of compactifications $\mathscr{C}$ in which
\begin{equation}\label{eq:modstab}
\text{dim}\,\Vac\bigl(\EFT(\mathscr{C})\bigr)=0\,,
\end{equation}
i.e.~in which the solutions to $\partial\Veff/\partial \phi^i=0$ are isolated points.
In such a case, the moduli
correspond to $d$-dimensional scalar fields with nonvanishing masses, which is a necessary step toward modeling the observed universe: astronomical and cosmological observations place strong limits on massless scalar fields.

More specifically, in moduli stabilization one usually begins with a well-understood reference compactification $\mathscr{C}_0$ that has a continuous moduli space: for example, this may preserve more than minimal supersymmetry.
One then seeks to make discrete but controllably small modifications $\mathscr{C}_0 \to \mathscr{C}$ such that
\begin{equation}\label{eq:cinc0}
    \Vac\bigl(\EFT(\mathscr{C})\bigr) \subset \Vac\bigl(\EFT(\mathscr{C}_0)\bigr)\,,
\end{equation}
with $\Veff(\mathscr{C})$ interpreted as
a nontrivial function on $\Vac\bigl(\EFT(\mathscr{C}_0)\bigr)$,
in order to achieve \eqref{eq:modstab}.

A canonical example discussed in detail in \S\ref{s:vacuumexample}
has $\mathscr{C}_0=(\IIbtheory,\K,G;\B{=}\varphi{=}\Lambda{=}w{=}0)$, with $K$ a Calabi-Yau threefold, $G$ the Ricci flat metric on $K$, and no branes or matter fields.
The task is then to identify consistent discrete choices of branes and fluxes $\B$ such that there is a related compactification
$\mathscr{C}=(\IIbtheory,\K',G',\B,\varphi,\Lambda<0,w)$, with $\K'$ an orientifold of $K$ (cf.~examples 9 and \ref{ex:IIbflux} of \S\ref{s:gen}), for which \eqref{eq:modstab} holds.  In such a case one says that the moduli of $\mathscr{C}_0$ have been stabilized by including fluxes.  Because $\Lambda<0$, the resulting vacua are AdS$_4$ vacua.

From the perspective of applications it would be equally interesting to find compactifications $\mathscr{C}$ obeying \eqref{eq:modstab} without starting from a reference compactification $\mathscr{C}_0$ with moduli, but it has generally proved easier to minimize a function $\Veff$ on a nontrivial space
$\Vac\bigl(\EFT(\mathscr{C}_0)\bigr)$ than to perform a fundamentally discrete search for isolated vacua.

Some care is required when comparing different points in the moduli space of an EFT.
One of the most important limitations is the
possibility of new states becoming light.
Consider an EFT $\Etheory$ and a path in $\Vac(\Etheory)$, parameterized as $\phi(t) \in \Vac(\Etheory)$ with $t\in[0,1]$.
Recall from \S\ref{sec:eft} that the definition of an EFT involves a cutoff $M$ and excludes fields with mass $m>M$.
Since the masses (squared) of the scalar fields in $\Etheory$ are
the eigenvalues $m_i^2$
of the Hessian $\delta^2 V_{\text{eff}}$, the number of scalar fields in $\Etheory$ at a point $\phi \in \Vac(\Etheory)$ is
\begin{equation}\label{eq:fieldcount}
    N(\phi,M) = \# \,\Bigl(\text{eig}\,\delta^2 V_{\text{eff}} < M^2\Bigr)\,.
\end{equation}
As we move in $\Vac(\Etheory)$, it can happen that $N(\phi,M)$ decreases as a field's mass rises above the cutoff, with $m(t{=}0)<M$ and $m(t{=}1)>M$.  In this case we can simply integrate out the offending
field, cf.~\eqref{eq:rgmap}.
However, the reverse is also possible: a field with $m(t{=}0)>M$, which was integrated out in defining the original EFT $\Etheory$, could have $m(t{=}1)<M$ and belong in the new EFT.  In this case we say that the EFT $\Etheory$ breaks down.  Because of this possibility,
the
global description of the moduli space of a compactification $\mathscr{C}$ may involve a collection of EFTs, each serving as a chart.

Even if an EFT has many vacua, in practice one often focuses on a particular vacuum $\phi_0$ and looks at its physics.
We denote the EFT $\theory$ at this vacuum as $\theory|_{\phi_0}$, and the EFTs in a small neighborhood of it as
$\theory|_{\phi\rightarrow\phi_0}$.  In particular, the weak string coupling regime is $\theory|_{g_s\rightarrow 0}$,
and the large volume limit is $\theory|_{\mathcal{V}\rightarrow \infty}$.

At weak coupling, small fluctuations of the fields correspond to particles, higher-order terms in the
effective action describe the interactions of the particles, and one can do QFT perturbation theory.  Mathematically, one is
writing $\phi=\phi_0+\delta\phi$ and doing an expansion in powers of $\delta\phi$.
Standard arguments equate the expansion in powers of the fluctuations
to an expansion in Planck's constant $\hbar$.   Although the EFT has an exact definition in the quantum theory,
very often all we know about it is series expansions in $\hbar$ around various vacua.
Later we will see other expansion
parameters for which this is the case.
We discuss the specifics further in \S \ref{ss:superQ}.

\subsection{Effective field theory -- derivation }

A compactification $\mathscr{C}$ gives rise to a $d$-dimensional effective field theory $\EFT(\mathscr{C})$
in a canonical way.  First, if $\theory$ includes $\GR$, the EFT will include $\GR$.  The Einstein term in the action is rewritten
\begin{eqnarray} \label{eq:Planck}
    M_{\text{pl},D}^{D-2} \int_{\CM_d} \int_K d^py\; \sqrt{g^{(D)}} R^{(D)} &=& M_{\text{pl},d}^{d-2} \int_{\CM_d} \sqrt{g^{(d)}} R^{(d)} + \dots \, ,
\end{eqnarray}
which relates the $d$-dimensional Planck scale $M_{\text{pl},d}$ to the $D$-dimensional Planck scale $M_{\text{pl},D}$, and using
$\sqrt{g^{(D)}}R^{(D)}=w^{2-4/d}\;\sqrt{g^{(\dimK)}}R^{(d)}$, to
the volume $\mbox{Vol}(K)$ and the warp factor $w$ of Eq.~(\ref{eq:ansatz}).  A choice of the units of length must then be made; usually
one chooses units with fixed $M_{\text{pl},d}$ (called ``Einstein frame'') and absorbs any moduli dependence of the other quantities, such as $\mbox{Vol}(K)$, into $w$.

Next, the data $(\K,G,\varphi,\B)$ determine the fields and action of the EFT.
In a few cases, the EFT is uniquely determined by symmetry: $\int_{T^k} \Mtheory$ is the maximal ungauged $d=11-k$ supergravity,
and $\int_{S^k} \Mtheory$ is the maximal gauged $d=11-k$ supergravity.
But in general, while symmetry leads to strong constraints, one must do geometric computations to derive the EFT.
Without going through these, let us explain some results and what they signify.

Consider example 1 of \S\ref{ss:examples}, with $\theory=\GR$.
The goal is to get an EFT whose solutions agree with those of $\int_K \GR$.
Let $\K\rightarrow F\xrightarrow{\pi} \Mod(K)$ be the family
of Ricci flat metrics on $\K$, parameterized by the moduli space $\Mod(K)$.
One can produce a large set of $D$-dimensional metrics as follows:
choose a function $\m:\CM_d\rightarrow \Mod(K)$, then the pullback $\m^* F$ is a $D$-dimensional space-time,
and the pullback metric has the form
of
Eq.~(\ref{eq:ansatz}) with the metric $G$ on $K$ depending on the location in $\CM_d$.
Taking $w=1$ and $A=0$, we get a $D$-dimensional metric $ds^2[\m]$, which generally no longer has maximal space-time symmetry.

Which of these metrics $ds^2[\m]$ are approximate solutions of the $D$-dimensional vacuum Einstein equations?
To simplify the problem, we suppose the function $\m$ is slowly varying, say with control parameter
$\epsilon\equiv\mbox{diam}(\K)||\partial\m|| \ll 1$.  Then
the metrics that are approximate solutions are those for which the fields $\m$ are classical solutions
of the EFT $\Seff\bigl(\Mod(K)\bigr)$, where the metric on $\Mod(K)$ is the Weil-Petersson metric.  More generally, the EFT
encodes the moduli spaces of the geometric objects in $\mathscr{C}$
(such as manifolds, connections, and branes) carrying natural metrics.

\bigskip
\noindent
Having introduced the ideas, let us sketch the derivation of the EFT.
We start with the $D$-dimensional action, obtain the $d$-dimensional Einstein action as in
Eq.~(\ref{eq:Planck}), and consider the remaining terms, dividing these into terms without and with $d$-dimensional derivatives.
The non-derivative terms contribute to the effective potential $\Veff$, while the two-derivative terms determine $\mathscr{T}$
with its metric, and $\mathcal{G}$ with its action on $\mathscr{T}$.\footnote{
Strictly speaking, $\mathcal{G}$ and $\mathscr{T}$ correspond to
the two-derivative terms of the form $g^{\mu\nu}\partial_\mu\ldots\partial_\nu\ldots$.  There can be other two-derivative
terms, as well as terms with more derivatives.
}  Then, we carry out the following steps:
\begin{enumerate}
    \item Define $\int_K \theory$ as an infinite-component field theory.
    \item Choose a background $G$ and $\varphi$ which solve the $D$-dimensional equations $\delta S=0$.
    Expand small fluctuations around the solution in a basis that diagonalizes the quadratic terms $\delta^2 S$,
    and reinterpret this expansion as the EFT expansion $\phi=\phi_0+\delta\phi$.
    \item Integrate out fields with mass $m>M$ to get a finite-component EFT with cutoff $M$.
    \item This gives a description in terms of patches around vacua $\phi_i$; given enough patches,
    in principle their overlaps and transition functions define the full EFT $\int_{K,B} \theory$.
\end{enumerate}
Let us flesh this out for the effective potential.
We define $\Veff(\delta\phi)$
by
substituting
the parameterized metrics $G(\phi_0+\delta\phi)$ and other fields $F(\phi_0+\delta\phi)$ into the $D$-dimensional action,
keeping the term without $d$-dimensional derivatives,
and factoring it into the integral over $\CM_d$ and the integral over $\K$,
\begin{equation} \label{eq:genVeff}
   \int_{\CM_d} \sqrt{g^{(d)}}\; \Veff(\delta\phi) := \int_{\CM_d}\int_K \sqrt{G}\, w^2 \cdot \Bigl( - R[G] + T[G,F]\Bigr)\bigg|_{(G,F)=(G,F)(\phi+\delta\phi)}\,,
\end{equation}
where $R[G]$ is the $\dimK$-dimensional Ricci scalar for the metric $G$,\footnote{The minus sign comes from the usual $S=\mbox{kinetic energy}-\mbox{potential energy}$. }
and $T[G,F]$ denotes the other terms.\footnote{
In addition to often appearing as a sum of squares, the energy $T$ due to matter satisfies the strong energy condition of GR and
is thus non-negative, with the crucial exception of certain string/M-theory correction terms whose presence allows Minkowski
vacua with nonzero matter configurations to exist.   An example is the term giving rise to Eq. (\ref{eq:bianchi}) below.
The stringy terms can be negative but have topological properties which imply that they satisfy {\it a priori} lower bounds.
}
In step 1, we interpret the right hand side as a functional of the metric $G$ and other fields $F$ on $\K$.
We then implement step 2 as sketched above, writing $G$ and $F$ as series expansions in $\delta\phi$ around a background solution,
substituting this expansion, and doing computations on $\K$ (multiplying normal modes and integrating) to derive explicit
terms in the left hand side $\Veff$.

However, even in step 1,
there is a problem.  Namely, physics normally requires a potential energy to be bounded below,
but Eq.~(\ref{eq:genVeff}) is not.  By making a local variation of the conformal factor $G\rightarrow e^f G$, it is possible
to make the scalar curvature $R[G]$ arbitrarily large and positive in a bounded region.  This is a very general feature of $\GR$, whose
resolution uses the fact that the $D$-dimensional variational equation
\begin{equation}\label{eq:constraint}
\delta S/\delta g_{00}=0
\end{equation}
is not an equation
of motion, but rather a constraint on the configuration space \cite{carlotto2021general}.
To implement this constraint in EFT, one can rewrite Eq.~\eqref{eq:constraint} in the form \cite{Douglas:2009zn}
\begin{eqnarray}\label{eq:warpcon}
    -\Delta w + \bigl(R[G] - T[G,F]\bigr) w = \lambda w^{1-4/d} ,
\end{eqnarray}
where $w$ is the warp factor of Eq.~(\ref{eq:ansatz}), and the constant $\lambda$ is determined
by fixing $M_{\text{pl},d}$ in Eq.~(\ref{eq:Planck}).  This equation has a unique solution $w$, which falls off exponentially in
regions with $R[G]\gg 0$, suppressing the negative contribution to the energy.   Reference \cite{Disconzi:2012jx} proved
that Eq.~(\ref{eq:genVeff}) with $w$ satisfying Eq.~(\ref{eq:warpcon}) is bounded below.

The first variation of Eq.~(\ref{eq:genVeff}), $\delta \Veff/\delta (G,F)=0$, with the constraint Eq.~(\ref{eq:warpcon}),
leads to geometric PDEs defining compactification as discussed in \S \ref{sec:rig}, whose solutions are allowed choices
of the background in step 2.  Given a solution, one proceeds to consider its small perturbations,
which satisfy the linearized equations of motion.  These are linear combinations of solutions $\delta\Phi_i$ of
$D$-dimensional
wave equations
\begin{equation}\label{eq:box}
\Box\delta\Phi + L\delta\Phi = 0\,,
\end{equation}
where $L$ is the linear operator on $\Phi$
obtained from taking second variations of Eq.~(\ref{eq:genVeff}).  The eigenvalues of $L$ are masses squared, denoted $m_i^2$.  In the EFT, the $m_i^2$ are the eigenvalues of the Hessian $\delta^2 \Veff$ evaluated in
an orthonormal basis for the metric on $\mathscr{T}$ at $\Phi_0$.
If any $m_i^2<0$, there will be
exponentially growing perturbations, called tachyons, and the solution
is unstable.\footnote{The stability criterion is modified for AdS compactifications.}
The remaining perturbations can be divided into {\term massless fields},
corresponding to {\term zero modes} of the linearized equations of
motion on $\K$, and {\term massive fields}, the others.  General
results on eigenvalues of Laplacians imply that the masses of massive
fields depend on the diameter of $\K$ as
$m \sim 1/{\rm diam}(\K)$, so at energies far smaller than $m$,
the massive fields cannot be excited.\footnote{
This restriction is not universal; given strong negative curvature on $K$, or a rapidly-varying warp factor,
one can have perturbations of small non-zero mass.}

To continue and get a physical EFT
which only depends on a finite number of $d$-dimensional fields, we must
carry out step 3, by integrating out the massive modes with $m_i>M$.
One can show that  for any fixed cutoff
$M$  there are a finite number of modes with $m_i\le M$.  Thus, one could eliminate all dependence on the massive modes by
solving their equations of motion and substituting these solutions into the action, to produce the EFT action.  This step is mathematically
well defined but rarely done explicitly; rather one normally sets the massive fields to zero and
appeals to general arguments that the error in doing this should be of order $1/M$.
A few compactifications have the property of consistent truncation, meaning that the error is zero
and this procedure is exact.

A supersymmetric theory $\theory$ admits supersymmetric compactifications.
While the discussion above still applies, now $\K$ and $\int_K \theory$ satisfy additional constraints,
as we discuss in \S \ref{s:susy}.
These constraints will enable us to get explicit expressions for $\Veff$, as we discuss in \S \ref{s:typeII}.

\section{Supersymmetric compactification}
\label{s:susy}

If $\theory$ is a supergravity and $\K$ admits a covariantly constant spinor, i.e.~a spinor $\epsilon$ obeying
\begin{equation}\label{eq:covcon}
\nabla_I\epsilon=0\,,
\end{equation}
then the compactified theory will also be
a supergravity theory.
The necessary condition is to have a local supersymmetry transformation $\epsilon$ which
preserves the gravitino $\psi_I$, schematically
\begin{equation}
    0 = \delta \psi_I = (\nabla_I+\varphi\cdot\Gamma_I) \epsilon .
\end{equation}
This is a highly overdetermined system of equations which generally admits no solutions.  A large class of
solutions are either highly
symmetric (for example, with $\K$ a sphere) or else are related to solutions of the simpler equation \eqref{eq:covcon},
say by multiplication by a scalar function.
A nonzero solution of \eqref{eq:covcon} defines a trivial component of the holonomy action on spinors on $\K$.  Thus, supersymmetric compactification is closely related to the theory of special holonomy.
Of particular note are the manifolds with $SU(n)$ holonomy, usually called Calabi-Yau $n$-folds or $\CalabiYau_n$.

Another important ingredient is the classification of string/M theories.
$\theory$ can be $D=11$ M-theory, denoted $\Mtheory$; a $D=10$ heterotic string with $E_8\times E_8$ or $Spin(32)/\BZ_2$ gauge
group ($\HEtheory$ and $\HOtheory$); a $D=10$ closed superstring theory ($\IIatheory$, $\IIbtheory$);
or one of these with orientifolds and branes ($\Itheory$, $\IIatheory'$ or $\IIbtheory'$).
These theories are related by superstring dualities: $\IIatheory=\int_{S^1} \Mtheory$, $\int_{S^1}\IIbtheory=\int_{T^2} \Mtheory$,
$\HEtheory=\int_{I^1} \Mtheory$ where $I^1\cong [0,R]$ is a real interval.
Mirror symmetry \cite{cox1999mirror} is a duality:
given a mirror pair of CY threefolds $M$ and $W$,  $\int_M \IIatheory=\int_W \IIbtheory$.
Another important duality is
$\int_{K3} \Mtheory = \int_{T^3} \HEtheory$.
F-theory is defined by $\int_K\Ftheory=\int_{K',\B}\IIbtheory$ where
(part of) the brane data $\B$ determines an elliptic fibration $E \rightarrow K \rightarrow K'$, and many of the constraints
on $\K'$ and $\B$ follow from taking $\K$ to be a manifold of $SU(n)$ holonomy.

Combining the choices of the holonomy group of $K$ with those of $\theory$, we get the options in Table 1, whose
entries indicate the space-time dimension $d$ and the number of supercharges
$\mathcal{N}\cdot s$.\footnote{
Here $s$ is the minimal number of supercharges in dimension $d$: $d=2,3,4,5,6,\ge 7$ have $s=1,2,4,8,8,16$ respectively.
In $d=2,6,10$ the supersymmetries come in two varieties, with left and right chirality, but the table only gives the combined number.
}
The observed case is of course $d=4$,
with $\mathcal{N}\cdot s=4$ or $0$ (otherwise one cannot get the chiral matter of the Standard Model).  Most physics work is on the five
middle entries of the table that realize this case:
$\HEtheory$, $\IIatheory'$, and $\IIbtheory'$ on $\CalabiYau_3$; $\Ftheory$ on $\CalabiYau_4$; and $\Mtheory$ on $G_2$.

\begin{table}[H]
$$\begin{array}{c|c|c|c|c|c|c|c}
& \text{torus} & \text{SU(2)} & \text{SU(3)} & \text{SU(4)} & \text{G}_2 & \text{Sp(4)} & \text{Spin(7)} \\
\hline
\text{\Mtheory,\IIatheory,\IIbtheory} & \text{any}/32 & 6^{+1}/16 & 4^{+1}/8 & 2^{+1}/4 & 3^{+1}/4 & 2^{+1}/6 & 2^{+1}/2 \\
\text{\Ftheory,\IIatheory$'$,\IIbtheory$'$} & \text{any}/16 & 6^{+2}/8 & 4^{+2}/4 & 2^{+2}/4 & 3/2 & 2/3 & 2/1 \\
\text{\HEtheory,\HOtheory,\Itheory} & \text{any}/16 & 6/8 & 4/4 & 2/2 & 3/2 & 2/3 & 2/1
\end{array}
$$
\noindent
Table 1 -- The rows are string/M theories, the columns are holonomy groups, and the entries are
the resulting dimension $d$ and total number of supercharges $\mathcal{N}\cdot s$.
The superscripts are a reminder that $d\rightarrow d+2$ in F-theory and $d\rightarrow d+1$ in M-theory.
For example, the entry $\text{\Mtheory,\IIatheory,\IIbtheory}: 2^{+1}/4$ in the $\text{SU(4)}$ column
means that $\IIatheory$ on a manifold of holonomy $\text{SU(4)}$
has $\mathcal{N}\cdot s=4, d=2, s=1$,
while $\Mtheory$ on such a manifold has $\mathcal{N}\cdot s=4, d=3, s=2$.
The theories
$\IIatheory'$ and $\IIbtheory'$ denote the superstring theories
$\IIatheory$ and $\IIbtheory$ with fluxes and branes incorporated, e.g.~via orientifolding.
\end{table}

\subsection{Rigorous work and existence results }\label{sec:rig}

While string/M theory is not a $D$-dimensional field theory, for some physical questions it can be well approximated by
a $D$-dimensional maximal supergravity theory with a few ``stringy'' corrections.  Making this approximation, each entry in table 1
gives rise to a system of geometric PDEs, which can be studied rigorously.  The basic example is $SU(n)$ holonomy
and $\GR$ (with matter fields set to zero, $\varphi=0$),
for which the metric $G$ must satisfy the vacuum Einstein equation, and thus must
be Ricci-flat K\"ahler.  The existence of these metrics and the structure of their
moduli space is determined by Yau's theorem.

One can reduce $\GR$ on a fibration (as in \S \ref{ss:examples}, example \ref{ex:fiber}) to get an EFT
$\int_{K_1} \GR \sim \GR + \Sigmatheory\bigl(\Isom(K_1),\Mod(K_1)\bigr)$.
This relates the $\Vol(K_1)\rightarrow 0$
limits of metrics on $K$ to solutions of $\int_{K_2,B} \int_{K_1} \GR$.  The case of $K_1=T^3$
is the starting point for the SYZ conjecture.
This relation has been made rigorous in some cases: see for example \cite{tosatti2010adiabatic}.

Moving beyond the Ricci-flat K\"ahler case,
another mathematically well-studied
class of compactifications are
the heterotic theories $\HEtheory$ and $\HOtheory$ with $K$ a Calabi-Yau manifold,
which gives rise to the Hull-Strominger system \cite{Hull:1986kz,Strominger:1986uh}.
This involves a conformally K\"ahler metric, a bundle $V$ with a Hermitian Yang-Mills connection of zero first Chern class,
the scalar dilaton and a three-form gauge field strength $H$.
A distinctive feature is that $H$ appears as torsion in the metric connection, and  also satisfies the modified Bianchi identity
\begin{equation} \label{eq:bianchi}
    dH = \alpha'\, \Tr\left( F \wedge F - R \wedge R \right) ,
\end{equation}
where $F$ and $R$ are the curvatures of the Yang-Mills and metric connections, respectively, and $\alpha'$ is a constant
with units of length squared.
The right hand side of \eqref{eq:bianchi} is a ``stringy correction'' in the sense of \S \ref{s:gen}.
The K\"ahler condition is relaxed to require a positive $(1,1)$-form $\omega$ such that $dH=\partial\bar\partial\omega$ and $d(||\Omega^{(3,0)}||\omega^2)=0$.

There are some necessary conditions for the existence of a solution to the Hull-Strominger system.
Eq.~(\ref{eq:bianchi}) forces the topological constraint $c_2(V)=c_2(K)$ (in the sense of $\partial\bar\partial$ cohomology),
but this must be satisfied with $dH\ne 0$ (so that $\omega>0$).
Furthermore $V$ must be slope stable to admit a Hermitian Yang-Mills connection.
Finally there must exist a positive closed $(1,1)$-form $\tilde\omega$.
Yau has conjectured that these are also sufficient conditions for existence \cite{Fu:2006vj};
recent work on this topic includes \cite{Garcia-Fernandez:2023nil}.

\subsection{General supersymmetric EFTs }
\label{ss:superEFT}

We now briefly review constraints on EFTs from supersymmetry.  The best-known case is $\mathcal{N}=1$ supersymmetry in $d=4$,
for which $\mathscr{T}$ must be a complex K\"ahler manifold, whose K\"ahler metric can be encoded in a K\"ahler potential $\mathcal{K}$.
The effective potential is derived from a holomorphic function\footnote{
In supergravity, $W$ is a section of a line bundle $\mathcal{L}$ on $\mathscr{T}$
with curvature $F=\partial{\bar\partial}\mathcal{K}$ (see the article on supergravity).
}
on $\mathscr{T}$,
the superpotential $W$, as
\begin{equation}\label{eq:vwk}
    \Veff = ||\nabla W||^2 - \frac{3}{M_{\text{pl},d}^2} ||W||^2\,,
\end{equation}
where $||\ldots||^2$ and $\nabla$ are covariant norms and derivatives for $\mathcal{L}$.  In \S\ref{sec:eft} we wrote the data of an EFT as $\Seff(\mathscr{G},\mathscr{T},\Veff)$, with a field space $\mathscr{T}$, isometries $\mathscr{G}$, and effective potential $\Veff$.  For an $\mathcal{N}=1$ supersymmetric EFT, these data can be replaced by
\begin{equation}\label{eq:sne1}
    \Seff_{\mathcal{N}=1} = \Seff(\mathcal{K},W)\,,
\end{equation}
with $\mathscr{T}$ and $\mathscr{G}$ obtained from $\mathcal{K}$,
i.e.~with the understanding that $\mathcal{K}=\mathcal{K}(\mathscr{T})$ denotes a K\"ahler potential for $\mathscr{T}$, and with $\Veff$ computed from $\mathcal{K}$ and $W$ using Eq.~\eqref{eq:vwk}.

Vacua can be supersymmetry-preserving or supersymmetry-breaking.
In global supersymmetry (the limit
$M_{\text{pl,d}}\rightarrow\infty$), the supersymmetry-preserving vacua are the quotient of the space of critical points $\partial W=0$
by $\mathcal{G}$ in the sense of symplectic quotient, meaning that the moment maps (D-terms) are set to zero.  The analogs
for broken supersymmetry and with finite $M_{\text{pl,d}}$ can be found in many references such as \cite{Denef_2005}.

For $\mathcal{N}=2$ supersymmetry in $d=4$, $\mathscr{T}$ is locally a product of two types of special geometries:
\begin{equation}\label{eq:tneq2}
\mathscr{T}_{\mathcal{N}=2} = \mathscr{T}_{\text{V}} \times \mathscr{T}_{\text{H}}\,,
\end{equation}
where  $\mathscr{T}_{\text{V}}$ is a so-called ``special K\"ahler geometry'' (parameterized by scalar fields in supersymmetry representations known as vector multiplets),
and $\mathscr{T}_{\text{H}}$ is
a
quaternionic K\"ahler geometry (parameterized by scalar fields in hypermultiplets).\footnote{Special K\"ahler manifolds are K\"ahler manifolds, but quaternionic K\"ahler manifolds are generally not K\"ahler manifolds.}
This still allows for an infinite set of possibilities, and the most basic classification uses the dimensions of the
two factors in $\mathscr{T}_{\mathcal{N}=2}$.  Many such theories can be obtained as compactifications from $\mathcal{N}=1$ in $d=6$, for which the
classification is finer; the vector multiplets of $d=4$ can come from either vector or tensor multiplets in $d=6$.

The cases $3\le \mathcal{N} \le 8$ in $d=4$ are far more highly constrained.
For maximal supersymmetry, $\mathcal{N}=8$, conjecturally there is
a finite list of  theories, and the corresponding compactifications
$\mathscr{C}$
are largely known.

\subsection{General remarks on the quantum theories }
\label{ss:superQ}

From a physics point of view, rigorous existence results as in \S \ref{sec:rig} are only a first step towards
establishing that the compactifications are solutions of quantum string/M theory.
Unfortunately, there is at present no mathematical definition of quantum string/M theory.  Most work on string compactification
follows the following approach:
\begin{itemize}
    \item Choose a $\theory$ and $\K$ from the table above, and choose a large structure limit of $\K$.  This could mean
    large or small $\vol(\K)$, or we could choose a fibration of $\K$ and take large or small volume for the base or fiber.
    \item Many large structure limits have a perturbative/semiclassical expansion.  For example,
    $\int_{S^1} \Mtheory|_{\text{Vol}(S^1)\rightarrow 0} = \IIatheory|_{g_s\rightarrow 0}$, for which one can use worldsheet perturbation theory.
    The parameter that plays the role of $\hbar$ is the string coupling $g_s = (M_{\text{pl},11}\Vol(S^1))^{3/2}$.
    \item Translating the $D$-dimensional moduli into moduli of $\int_K T$, the large structure  limits correspond to limits in the EFT moduli space around which
    there is a systematic expansion procedure.  This includes Feynman diagrams, leading to powers of $\hbar$; semiclassical computations of
    the effect of instanton solutions, which are generally $\exp(-c/\hbar)$; or both.
\end{itemize}
This approach is not completely satisfying: the expansions are only asymptotic, the computations involved are very
complicated, {\it etc.}.  In some cases there are alternatives -- most importantly, when there is a duality that relates
the EFT or some part of it to a quantum
field theory.

The strongest results for quantum EFT can be obtained for holomorphic quantities \cite{Seiberg:1995qzw}:
the superpotential for $\mathcal{N}=1$ supersymmetry in $d=4$,
and the prepotential for $\mathcal{N}=2$ in $d=4$ (this determines the moduli space metric).
The superpotential $W$ obeys nonrenormalization theorems.  These are usually stated for a field theory defined by quantizing
    an action: in this case $W$ receives no quantum corrections at any power of $\hbar$.  However, $W$ can receive exponentially small corrections,
    which usually can be attributed to specific instanton solutions.  There are also string theory versions based on world-sheet perturbation theory.
    By contrast, the K\"ahler potential $\mathcal{K}$ is ``unprotected'' and the quantum corrections satisfy no (known) general constraints.

\section{Type IIB compactification on Calabi-Yau orientifolds }
\label{s:typeII}

We will now illustrate the general ideas presented in \S\ref{s:gen}-\S\ref{s:susy} in a particular class of solutions: compactifications of type IIB string theory on orientifolds of Calabi-Yau threefolds.
For physical and mathematical reasons that we will explain, such compactifications have proved particularly fruitful in efforts to construct vacua of string theory with realistic
cosmology.
\subsection{Compactifications of type IIB supergravity}\label{sec:iibsugra}

We have repeatedly alluded to an approach to compactification of string theory in which one begins with a compactification of the corresponding supergravity theory, and then incorporates specific stringy corrections.  To apply this method to compactifications of type IIB string theory, denoted $\IIbtheory$ above, we will now introduce type IIB supergravity, find an initial vacuum configuration for the supergravity fields, and incorporate corrections from couplings to D-branes and orientifolds, and from stringy contributions to the effective action.

Type IIB supergravity is a ten-dimensional theory of gravitation and matter fields, whose bosonic\footnote{The fermionic fields, which we will not discuss explicitly, include two spin-$3/2$ gravitino fields of the same chirality.} fields are the ten-dimensional metric $g^{(10)}_{MN}$; a real scalar $\phi$, called the dilaton; and $p$-form potentials denoted $C_0$, $C_2$, $B_2$, $C_4$, with corresponding $(p+1)$-form field strengths, called fluxes: $F_1=dC_0$, $F_3=dC_2$, $H_3=dB_2$, and $F_5=dC_4$.
From these fields one defines the combination
\begin{equation}
    \tilde{F}_5 := F_5 + \frac{1}{2}B_2 \wedge F_3 - \frac{1}{2}C_2 \wedge H_3\,,
\end{equation} and
a complex scalar called the axiodilaton,
\begin{equation}
\tau := C_0 + i e^{-\phi}\,,
\end{equation} in terms of which we further define the complex three-form flux
\begin{equation}
G_3 := F_3 -\tau H_3\,.
\end{equation}
The action for the bosonic fields then reads
\begin{equation}\label{eq:siib}
    S = \frac{1}{2\kappa_{10}^2}\int \text{d}^{10}x\,\sqrt{-\text{det}\,g^{(10)}}\Biggl(R-\frac{\partial_{\mu}\tau\partial_{\nu}\bar{\tau}G^{\mu\nu}}{2\,(\text{Im}\,\tau)^2}-\frac{G_3\cdot\overline{G}_3}{2\,\text{Im}\,\tau}-\frac{\tilde{F}_5^2}{4}\Biggr)+\frac{1}{8i\kappa_{10}^2}\int \frac{C_4 \wedge G_3 \wedge \overline{G}_3}{\text{Im}\,\tau}\,,
\end{equation} where $\kappa_{10}$ is the ten-dimensional gravitational coupling, and contractions are performed with the metric $g^{(10)}_{MN}$.

The equations of motion for the bosonic fields, in the absence of stringy corrections, are the Euler-Lagrange equations resulting from \eqref{eq:siib}, supplemented by the self-duality constraint $\star \tilde{F}_5 = \tilde{F}_5$.  The configuration must also obey the
Bianchi identity,
\begin{equation}\label{eq:f5}
\text{d}\tilde{F}_5 = H_3 \wedge F_3 + \text{local}
\end{equation}
where the local source term receives contributions from D3-branes, anti-D3-branes, D7-branes, and orientifold planes: see e.g.~\cite{Kachru:2019dvo} for complete expressions for the source terms.

A trivial solution is ten-dimensional flat space-time, $\mathbb{R}^{9,1}$, with $\tau$  constant; $F_3$, $H_3$, and $F_5$ all vanishing; and no D-branes or orientifold planes, so that the local source term
in Eq.~\eqref{eq:f5} vanishes.

A more interesting  solution is a Calabi-Yau vacuum configuration, in which the internal space $K$ is a Calabi-Yau threefold.
In this case the warp factor $w$ in the metric ansatz \eqref{eq:ansatz} can be set to unity, the noncompact spacetime is Minkowski spacetime, and the mixed metric components $A_{\mu i}$ vanish, so that
\begin{equation}
    ds^2 = \eta_{\mu\nu}dx^\mu dx^\nu + G_{ij} dy^i dy^j\,,
\end{equation} with $G_{ij}$ the Ricci-flat metric on $K$.
Such a solution preserves $\mathcal{N}=2$ supersymmetry in four dimensions, and leads to an EFT that is strongly constrained as a result (cf.~\S\ref{ss:superEFT}).  Given a Calabi-Yau threefold $K$ with Hodge numbers $h^{1,1}$ and $h^{2,1}$, a vacuum configuration of type IIB supergravity on $K$ has a moduli space of (complexified) Ricci-flat metrics,
\begin{equation}\label{eq:mn2}
    \mathcal{M}_{\mathcal{N}=2} = \mathcal{M}_{\text{V}} \times \mathcal{M}_{\text{H}},
\end{equation}
with $\text{dim}_{\mathbb{C}}(\mathcal{M}_{\text{V}})=h^{2,1}$ and $\text{dim}_{\mathbb{R}}(\mathcal{M}_{\text{H}})=4h^{1,1}+4$.  The vector multiplet moduli space $\mathcal{M}_{\text{V}}$ is a special K\"ahler manifold parameterized by the possible complex structures of Ricci-flat metrics on $X$.
The hypermultiplet moduli space
$\mathcal{M}_{\text{H}}$ is a quaternionic K\"ahler manifold parameterized by the possible K\"ahler forms of Ricci-flat metrics on $X$; by the integrals of the $p$-forms $B_2$, $C_2$, $C_4$ on $p$-cycles in $X$; and finally by the axiodilaton $\tau$ and a further complex scalar obtained by dualizing $B_{\mu\nu}$ and $C_{\mu\nu}$ in four dimensions.\footnote{
Mirror symmetry relates $\mathcal{M}_{\text{H}}$ for compactification on a $\CalabiYau_3$, $M$, to $\mathcal{M}_{\text{V}}$ for compactification on its mirror $W$.
This implies that some apparent singularities in $\mathcal{M}_{\text{H}}$ (for example a limit in which the volume of a cycle would vanish)
are actually not singular in string theory, and one can continue through the apparent singularity to obtain a topologically distinct manifold $M'$.
This is an example of manifolds that differ in conventional topology, but are continuously connected in stringy topology.}

The high degree of unbroken supersymmetry guarantees that the potential for the moduli vanishes exactly.  Thus, the massless bosonic fields in four dimensions are the graviton and $2 h^{2,1} + 4h^{1,1} + 4$ real scalars.

Although the constraints of $\mathcal{N}=2$ supersymmetry are valuable in understanding the vacuum structure, they are so strong as to preclude any  modeling of realistic four-dimensional physics: the moduli are exactly massless, which is fatal for cosmology, and fermions cannot transform in chiral representations of gauge groups, and so one cannot accommodate the Standard Model.

In seeking compactifications that can be candidate ultraviolet completions for the EFT of the observed universe, one productive strategy is to begin with vacuum solutions preserving $\mathcal{N}=2$ supersymmetry, then introduce source terms that are a small perturbation, and so obtain a nearby  solution preserving $\mathcal{N}=1$ supersymmetry.  Because of the reduced supersymmetry, such a   solution has richer possibilities than an $\mathcal{N}=2$ Calabi-Yau vacuum configuration: in particular, moduli masses and chiral gauge representations are possible.  Next, one can attempt a further breaking to $\mathcal{N}=0$ supersymmetry, again by a controllably small source.

The above method is an example of the general approach indicated in Eq.~\eqref{eq:cinc0},
with the starting point $\mathscr{C}_0$ corresponding to the Calabi-Yau vacuum configuration without fluxes, so that $\Vac\bigl(\EFT(\mathscr{C}_0)\bigr)$ contains the well-understood $\mathcal{N}=2$ moduli space $\mathcal{M}_{\mathcal{N}=2}$, as in Eq.~\eqref{eq:mn2}, and $\Veff(\mathscr{C})$ is a function on
$\Vac\bigl(\EFT(\mathscr{C}_0)\bigr)$.

As we explained in \S\ref{ss:moduli},
fluxes generally introduce metric-dependent terms in $\Veff$,
cf.~Eq.~\eqref{eq:fluxe}.
In $\IIbtheory$, the three-form fluxes $F_3 = dC_2$ and $H_3=dB_2$ are indispensable source terms in the construction of solutions with stabilized moduli based on Calabi-Yau vacuum configurations, and   the resulting solutions are called \emph{flux compactifications}.

\subsection{The data of a compactification}

We will now explain the data characterizing a flux compactification of $\IIbtheory$, beginning at the level of an $\mathcal{N}=2$ supersymmetric Calabi-Yau vacuum configuration; then introducing an orientifold involution, its associated source terms, and fluxes; and finally (in \S\ref{ss:wnp}) incorporating stringy corrections to the effective action.

The first step is to specify the topological type of a Calabi-Yau threefold.
By Wall's theorem, the homeomorphism and diffeomorphism class of a simply-connected, torsion-free Calabi-Yau threefold is specified by its Hodge numbers, triple intersection numbers, and second Chern class.  Thus, a consistent choice of these ``Wall data'' fully determines the topological type.  A given topological equivalence class $\mathcal{E}$ may have many representatives $X, X' \in \mathcal{E}$ that are realized in different ways, e.g.~as subvarieties in different ambient spaces.

To proceed to a non-vacuum solution preserving $\mathcal{N}=1$ supersymmetry, one can find a holomorphic involution $\sigma: X \to X$ that acts on the holomorphic $(3,0)$ form $\Omega$ of $X$ as $\sigma(\Omega)=-\Omega$.  Projecting onto configurations invariant under $\Omega\circ \Omega_{\text{ws}}$, with $\Omega_{\text{ws}}$
a suitable orientation-reversing action on the string worldsheet, defines an orientifold of $X$, as noted in
\S\ref{ss:examples}.

The fixed loci, which fill the noncompact spacetime and are either points or four-cycles in $X$, thus extending over 3 or 7 spatial dimensions in total, are called O3-planes and O7-planes, respectively.  Orientifold planes are localized sources in the internal space: they contribute as negative-tension sources of stress-energy in the Einstein equations, and as negative-charge sources in the $\tilde{F}_5$ Bianchi identity \eqref{eq:f5}.

In \S\ref{s:gen} we explained that the data of a compactification is
$\mathscr{C}=(\theory,K,\B,G,\varphi,\Lambda,w)$, but thus far most of our discussion has involved $\theory,K,G$.  We can now be more specific about the remaining data.
Fix $\theory=\IIbtheory$ and take $X$ to be a member of a topological equivalence class $\mathcal{E}$ of Calabi-Yau threefolds.
We then specify a flux compactification in terms of the following data:
\begin{itemize}
        \item A holomorphic involution $\sigma: X \to X$  with $\sigma(\Omega)=-\Omega$, defining an orientifold projection $\Gamma'$.  The compactification manifold $K$ is then the orientifold $K=X/\Gamma'$.  The fixed loci are O3-planes and O7-planes, with the latter in some effective divisor classes $\O7 \in H_4(K,\mathbb{Z})$.
        \item{The topological data --- such as Chern classes and Wilson lines --- of the gauge bundles on D7-branes wrapping four-cycles $D$ in the classes $\O7$.}
        \item{A choice of two classes $F_3$, $H_3 \in H^3(K,\mathbb{Z})$, called quantized three-form fluxes.}
    \end{itemize}

At this stage, all the topological data $\mathscr{C}_{\text{top}}$ of $\mathscr{C}$ are determined.  The metric $G$, warp factor $w$, matter field configuration (in particular, the profile of the axiodilaton $\tau$ in $K$) and cosmological constant $\Lambda$ can in principle be computed\footnote{The solutions are generally not unique: for example, for a given $\mathscr{C}_{\text{top}}$ one may find solutions
with $\Lambda<0$ and with $\Lambda>0$.} by solving the equations of motion of $\EFT(\mathscr{C})$.
Let us therefore turn to examining $\EFT(\mathscr{C})$, by first developing an appropriate ansatz (\S\ref{s:ccy}) and then obtaining the $\mathcal{N}=1$ supersymmetric EFT (\S\ref{ss:wnp}).

\subsection{Conformally Calabi-Yau flux compactifications}\label{s:ccy}

We consider an ansatz that is a slight variation of Eq.~\eqref{eq:ansatz}:
\begin{equation}
ds^2 = e^{-6u(x)+2A(y)}\eta_{\mu\nu}dx^\mu dx^\nu
+e^{2u(x)-2A(y)} g_{mn}dy^m dy^n\,,
\end{equation}
where $u(x)$ is a ``breathing mode'' parameterizing the overall volume of $K$, $\text{Vol}(K) \propto e^{6u}$, and $e^{A}$ is a warp factor.
We write the five-form as
\begin{equation}
    \tilde{F}_5 = \bigl(1+\star_{10}\bigr)\,e^{-12u(x)}\sqrt{-\text{det}\,g}\,\text{d}\alpha(y)\wedge \text{d}x^0\wedge \text{d}x^1\wedge \text{d}x^2\wedge \text{d}x^3\,,
\end{equation}
with $\alpha(y)$ a real-valued function on $K$, and
we define the functions
\begin{equation}
    \Phi_{\pm}(y) := e^{4A} \pm \alpha\,,
\end{equation} and write the three-form $G_3$ as
\begin{equation}
    G_{\pm} := G_3 \mp i \star_{6} G_3\,,
\end{equation} with $\star_{6}$ the Hodge star on $K$.

In a general configuration of nontrivial sources (fluxes, D-branes, and orientifold planes),
Einstein's equations for the metric $g_{mn}$ on $K$ do not allow $g_{mn}$ to be Ricci flat.
However, an important observation made in Giddings et al.~(2001) is that for a certain restricted but nontrivial class of sources, $g_{mn}$ is the Ricci flat K\"ahler metric on $K$.

Specifically, if $\Phi_-=0$ and $G_-=0$, and if the only localized sources are O3-planes, O7-planes, D7-branes, and D3-branes\footnote{Incorporating the O3-planes, O7-planes, D7-branes, $F_3$, and $H_3$ as contributions to the right hand side of Eq.~\eqref{eq:f5}, one can compute the number $N_{D3}$ of D3-branes that is necessary for consistency.} (but not, for example, anti-D3-branes), then $g_{mn}$ is the Ricci flat K\"ahler metric and the axiodilaton $\tau$ is constant.
Because the warp factor $e^{A}$ is in general nontrivial, such solutions are called \emph{conformally Calabi-Yau}.
The supersymmetric EFT of such compactifications is comparatively well-understood, as we will now explain.

\subsection{The $\mathcal{N}=1$ effective theory}\label{ss:wnp}

Focusing on the fields of spin zero, and on terms in the Lagrangian involving at most two derivatives with respect to the coordinates in the four-dimensional spacetime, the data of
an $\mathcal{N}=1$ supersymmetric EFT
can be expressed in terms of a K\"ahler potential $\mathcal{K}$ and a superpotential $W$, cf.~Eq.~\eqref{eq:sne1}.

As explained in \S\ref{ss:superEFT}, the superpotential does not receive polynomial corrections in perturbation theory,
and
so the full superpotential is the sum of a classical term and a series of nonperturbative contributions,
\begin{equation}
W = W_{\text{classical}} + W_{\text{np}}\,.
\end{equation}
The classical superpotential involves the three-form flux $G$ via
\begin{equation}
W_{\text{classical}} = \int G_3 \wedge \Omega\,,
\end{equation}
where $\Omega$ is the $(3,0)$ form on $X$, and we have adopted a convenient normalization.

The nonperturbative superpotential includes a sum of contributions associated to effective divisors on $K$ with specific topological properties.  In particular, if $D$ is a smooth effective divisor on $K$ and we define the $\sigma$-graded sheaf cohomology groups $H^{i}_{\pm}(D,\mathcal{O}_D)$, $i=0,1,2$, with corresponding dimensions $h^{i}_{\pm}(D,\mathcal{O}_D)$, then $D$ supports a nonperturbative superpotential term if
\begin{equation}\label{eq:rigid}
h^{\bullet}_{+}(D,\mathcal{O}_D)=(1,0,0)\text{~~~~~and~~~~~}h^{\bullet}_{-}(D,\mathcal{O}_D)=(0,0,0)\,.
\end{equation}
Such an effective divisor is called rigid.

The magnitude of such a nonperturbative term is determined, in the semiclassical approximation, by the calibrated volume of $D$ obtained from the K\"ahler form $J$ of the Ricci-flat metric on $X$, with $\text{Re}(S) = \frac{1}{2}\int_{D} J \wedge J$.
We therefore write
\begin{equation}
    W_{\text{ED3}|D} = \text{exp}\biggl(-\int_D \ED3 \biggr)\,,
\end{equation}
with
\begin{equation}\label{eq:ed3j}
   \frac{1}{2\pi}\,\ED3 =  \frac{1}{2} J \wedge J + i C_4 - \vartheta\,,
\end{equation} and with
\begin{equation}
   \int_D\vartheta = \text{log}\Bigl(\theta_D\bigl(\Mod(K,B)\bigr)\Bigr)\,,
\end{equation}
with $\theta_D$ denoting a section of a certain holomorphic bundle on $\Mod(K,B)$.
The `Pfaffian'
\begin{equation}\label{eq:pfa}
    \mathcal{A}_D = \theta_D\bigl(\Mod(K,B)\bigr)\,,
\end{equation} encodes corrections to the semiclassical approximation.
The appearance of $J$ in Eq.~\eqref{eq:ed3j} indicates that $W_{\text{ED3}}$ gives rise to a potential on the moduli space of K\"ahler structures, with
\begin{equation}
    \bigl|W_{\text{ED3}|D}\bigr| = |\mathcal{A}_D| \,e^{-\text{Vol}(D)}\,.
\end{equation}

D7-branes wrapping a four-cycle $D$ support a (7+1)-dimensional gauge theory with gauge algebra $\mathfrak{g}_D$.  A single D7-brane on $D$ yields $\mathfrak{g}_D=\mathfrak{u}(1)$, while for multiple D7-branes one has a Yang-Mills theory, which we denote $\YM_{\mathfrak{g}_{D}}^{(8)}$, with $\mathfrak{g}_D$ determined by the configuration of O7-planes, if any, on $D$.  For example, if an O7-plane and 4 D7-branes coincide on $D$, then $\mathfrak{g}_D=\mathfrak{so}(8)$.

The EFT then contains four-dimensional Yang-Mills fields,
$\YM_{\mathfrak{g}_D}^{(4)}=\int_K \YM_{\mathfrak{g}_D}^{(8)}$.
Strong gauge dynamics at low energies (specifically, ``gaugino condensation'', denoted $\langle\lambda\lambda\rangle$) in such a theory
can give a further contribution to $W$.
Because the volume of $D$ determines the gauge coupling of $\YM_{\mathfrak{g}_D}^{(4)}$, the gaugino condensate superpotential can be expressed in terms of the same complexified volume, Eq.~\eqref{eq:ed3j}, that appears in $W_{\text{ED3}|D}$: we have
\begin{equation}
    W_{\langle\lambda\lambda\rangle,D}   = \text{exp}\Bigl(-\int_D \ED3/c_2(\mathfrak{g}_D) \Bigr)\,,
\end{equation} where $c_2(\mathfrak{g}_D)$ is the dual Coxeter number of $\mathfrak{g}_D$.

There is in addition a contribution from Euclidean D(-1)-branes, of the form
\begin{equation}
    W_{\Ed(-1)} = \sum_{n \in \mathbb{Z}}B_k\,\text{exp}\Bigl(-2\pi i n \tau\Bigr)\,,
\end{equation}
with some numerical prefactors $B_k$.
At sufficiently weak coupling, i.e.~for $\IIbtheory|_{g_s\to 0}$, the Euclidean D(-1)-brane superpotential $W_{\Ed(-1)}$ is negligible in comparison to the other terms.  We will make this approximation henceforth.

In total the superpotential is\footnote{The interplay between D7-brane and Euclidean D3-brane contributions can be intricate, and depends on how these branes intersect.
The form given in Eq.~\eqref{eq:wfull} suffices for the example of \S\ref{s:vacuumexample}.}
\begin{equation}\label{eq:wfull}
    W=\int_{K}   G_3 \wedge \Omega + \sum_{D \in H_4(K)} \text{exp}\biggl(-\int_{D}\ED3\biggr)+ \sum_{D \in \O7}\text{exp}\biggl(-\int_D \ED3/c_2(\mathfrak{g}_D) \biggr)\,,
\end{equation}
where the Euclidean D3-brane sum $\sum_{D \in H_4(K)}$ runs over effective divisors in $K$, with the Pfaffians from Eq.~\eqref{eq:pfa} enforcing a vanishing contribution for divisors $D$ for which the Dirac operator has too many zero modes.
The gaugino condensate sum  $\sum_{D \in \O7}$ only runs over effective divisors containing O7-planes, because of the D7-brane tadpole cancellation condition.

With the understanding that for $D \in \O7$, $\ED3 \to \ED3/c_2(\mathfrak{g}_D)$, we can write the total superpotential more concisely as
\begin{equation}\label{eq:wfullnew}
    W=\int_{K}   G_3 \wedge \Omega + \sum_{D \in H_4(K)} \text{exp}\biggl(-\int_{D}\ED3\biggr) \,.
\end{equation}

Thus far we have explained the contributions of effective divisors $D$ that fulfill the rigidity condition \eqref{eq:rigid}, and with no further conditions on $D$, the Pfaffian generally depends nontrivially on the complex structure moduli, axiodilaton, and the moduli governing seven-brane and D3-brane positions, i.e.~on all of $\Mod(K,B)$ except the K\"ahler moduli.  However, in cases meeting further topological conditions, one finds that $\mathcal{A}_D$ depends only on the positions of D3-branes, and so in compactifications without space-filling D3-branes, such a Pfaffian $\mathcal{A}_D$ is just a complex number.

Recalling Eq.~\eqref{eq:sne1}, and denoting by $\mathscr{T}$ the the field space
\begin{equation}
\mathscr{T} = \Mod(\K,\D3, \MD7)\,,
\end{equation}
parameterizing the geometric moduli of $K$, the axiodilaton, and the moduli of D3-branes and D7-branes,
and writing $\mathcal{K}(\mathscr{T})$ for the corresponding K\"ahler potential,
we can write the full EFT of a flux compactification as
\begin{equation}
\int_K \IIbtheory = \GR + \mathcal{S}\Biggl(\mathcal{K}\Bigl(\Mod(\K,\D3, \MD7)\Bigr),W=\int_{K} G_3 \wedge   \Omega + \sum_{D \in H_4(K)} \text{exp}\Bigl(-\int_{D}\ED3\Bigr)\Biggr)\,.
\end{equation}

\subsection{An example of a vacuum}\label{s:vacuumexample}

To make the considerations discussed above more concrete, we will give an example of a compactification in which the geometric data is explicit and the vacuum structure is well-understood.\footnote{This undertaking requires more detailed background, for which we refer the reader to \cite{Demirtas:2021nlu} and references therein.}

To define a compactification manifold,
we first construct a toric
variety $V$
of complex dimension four.
We begin with the reflexive polytope
$\Delta \subset \mathbb{Z}^4$, selected from the complete classification by Kreuzer and Skarke,
whose vertices are the columns of
\begin{equation}\label{eq:Delta}
\begin{pmatrix}
1  & -3 & -3 & 0  & 0  & 0 &-5 &-2  \\
0 & -2  & -1 & 0  & 0  & 1 &-3 &-1  \\
0 & \phantom{-}0   & -1 & 0  & 1  & 0 & \phantom{-}0 & \phantom{-}1  \\
0 & \phantom{-}0   & 0  & 1  & 0  & 0 &-1 &-1
\end{pmatrix}\, ,
\end{equation}
and we denote by $\Delta^{\circ}$ the polar dual of $\Delta$.
A suitable (specifically: regular and star, and fine with respect to points in two-faces) triangulation $\mathcal{T}$ of $\Delta^{\circ}$ defines a toric fourfold $V$ in which the generic anticanonical hypersurface is a smooth Calabi-Yau threefold, $X$, with a smooth mirror $\widetilde{X}$.  One finds that $h^{2,1}(X)=5$ and $h^{1,1}(X)=113$.
At this stage we have specified $X$ as an algebraic variety.  In the terminology of \S\ref{s:gen}, we have chosen $\theory=\IIbtheory$ and $K=X$, with $G$ the Ricci-flat K\"ahler metric\footnote{By choosing a section of the anticanonical bundle we have selected a point in complex structure moduli space, while thus far no point has been specified in the K\"ahler moduli space.} on $X$, with $\B$, $\varphi$, and $w$ trivial, and with $\Lambda=0$.

By computing the periods of the $(3,0)$ form $\Omega$ on $X$ in the limit of large complex structure (LCS) and using mirror symmetry, one can compute the Gopakumar-Vafa invariants of curves in the mirror threefold $\widetilde{X}$.  Similarly, by computing the periods of the $(3,0)$ form on $\widetilde{X}$, one obtains the Gopakumar-Vafa invariants of curves in $X$. In this way one can obtain the prepotential $\mathcal{F}$ --- and hence the K\"ahler potential $\mathcal{K}$ and classical superpotential $W_{\text{classical}}$ --- for the complex structure moduli space of type IIB string theory compactified on $X$, expanded around large complex structure.
One similarly obtains information about the metric on K\"ahler moduli space, expanded around large volume.  That is, by performing period integrals one obtains the asymptotic approximations
\begin{equation}   \mathcal{K}\bigl(\mathcal{M}_{\text{V}}\bigr)\Bigl|_{\text{LCS}}\,,
\end{equation}
\begin{equation} \label{eq:wclper}  W_{\text{classical}}\bigl(\mathcal{M}_{\text{V}}\bigr)\Bigl|_{\text{LCS}}\,,
\end{equation}
and
\begin{equation}
\mathcal{K}\bigl(\mathcal{M}_{\text{H}}\bigr)\Bigl|_{\mathcal{V}\to \infty}\,.
\end{equation}
In the case of the classical superpotential, Eq.~\eqref{eq:wclper}, one can in principle obtain an arbitrarily good approximation.

The toric variety $V$ admits an involution $\sigma_V$ corresponding to a sign flip of a particular toric coordinate $x_1$,
\begin{equation}\label{eq:oact}
    \sigma_{V}: x_1 \to -x_1\,,
\end{equation}
and this induces an orientifold involution on $X$.
None of the geometric moduli are projected out by this orientifold action.  The fixed loci of the orientifold are 26 O7-planes, each wrapping a prime toric divisor, and 48 O3-planes at the intersections of certain triples of prime toric divisors.

There exist choices of quantized three-form  fluxes $F_3$, $H_3 \in H^3(X,\mathbb{Z}) = \mathbb{Z}^{2h^{2,1}+2}$
that cause the complex structure moduli and axiodilaton to be stabilized in an isolated minimum $z^a_{\star}$, $\tau_{\star}$ where the expectation value of the flux superpotential is small.
For example, the fluxes
\setcounter{MaxMatrixCols}{20}
\begin{align}\label{eq:fluxchoice}
&\vec{f}=\begin{pmatrix}
10&  12 & 8 & 0 & 0 & 4 & 0 & 0 & 2 & 4 & 11 & -8
\end{pmatrix}\, ,\nonumber\\
&\vec{h}=\begin{pmatrix}
0 &  8 & -15 & 11 & -2 & 13 & 0 & 0 & 0 & 0 & 0 & 0
\end{pmatrix}\, ,
\end{align}
lead to the effective single-field classical superpotential
\begin{equation}
   W_{\text{classical}} =  W_{\text{eff}}(\tau) = \sqrt{\tfrac{2^3}{\pi^5}} \Biggl(-2\,e^{2\pi i \tau\cdot\frac{7}{29}} + 252\,e^{2\pi i \tau\cdot\frac{7}{28}}\Biggr)  + \mathcal{O}\Bigl(e^{2\pi i \tau\cdot\frac{43}{116}}\Bigr) \,,
\end{equation}
which is minimized at the value
\begin{equation}\label{eq:w0value}
W_0 \equiv \bigl\langle \bigl|W_{\text{eff}}\bigr| \bigr\rangle_{\tau^{\star}} \approx \biggl(\frac{2}{252}\biggr)^{29}\,.
\end{equation}
Integrating the Bianchi identity \eqref{eq:f5} over $K$ and using the flux choice \eqref{eq:fluxchoice}, one finds
\begin{equation}
0 = \int_X H_3 \wedge F_3 + \int_X \text{local}_{\text{O3/O7}} + \int_X \text{local}_{\text{D3}} = 56 - 60 + N_{\text{D3}}\,,
\end{equation}
where $\int_X \text{local}_{\text{O3/O7}}=-60$ is the contribution of the orientifold planes, and $N_{\text{D3}}$ is the number of spacetime-filling D3-branes.
Thus we must take $N_{\text{D3}}=4$, i.e.~four spacetime-filling D3-branes need to be included for
the $\tilde{F}_5$ Bianchi identity \eqref{eq:f5} to be consistent.  This is known as a ``tadpole constraint''.

The Hessian for the complex structure moduli and the axiodilaton at the point $z^a_{\star}$, $\tau_{\star}$ is positive-definite, but the classical effects described so far leave the potential for the K\"ahler moduli flat.

To understand the nonperturbative potential for the K\"ahler moduli,
we first examine the 26 O7-plane fixed loci of the orientifold action Eq.~\eqref{eq:oact}.
One of the fixed prime toric divisors (call it $D_0$) has
$h^{\bullet}_{+}(D_0,\mathcal{O}_{D_0})=(1,0,2)$
and supports an $\mathcal{N}=1$ Yang-Mills theory with gauge algebra $\mathfrak{so}(8)$ and two chiral multiplets transforming in the adjoint representation.
This cycle will turn out to give a negligible contribution to the
scalar potential of the
EFT at $t^{\star}$.
The remaining 25
fixed prime toric divisors $D_1,\ldots, D_{25}$ are rigid, with
$h^{\bullet}_{+}(D,\mathcal{O}_D)=(1,0,0)$,
and each supports an $\mathcal{N}=1$ Yang-Mills theory with gauge algebra $\mathfrak{so}(8)$ and no matter fields.
The topology of $D_1,\ldots, D_{25}$ in this example is such that in each case the Pfaffian $\mathcal{A}_{D_i}$
is independent of the complex structure moduli and axiodilaton,\footnote{In the example presented here, the Pfaffians depend only on the positions $z_{D3}$ of the  spacetime-filling D3-branes, which we suppress here for simplicity.  Details are given in \cite{Demirtas:2021nlu}.}
and we have, using $c_2(\mathfrak{so}(8))=6$,
\begin{equation}\label{eq:wll}
    W_{\langle\lambda\lambda\rangle} = \sum_{i=1}^{25}\text{exp}\biggl(-\frac{1}{6}\int_{D_i} \ED3  \biggr)\,.
\end{equation}
Moreover,
one
finds 89 Euclidean D3-brane contributions from rigid prime toric divisors $D_{26},\ldots, D_{114}$,
where again the Pfaffians depend only on the D3-brane positions:
\begin{equation} \label{eq:wed3}
W_{\text{ED3}} = \sum_{i=26}^{114} \text{exp}\biggl(-\int_{D_i} \ED3 \biggr)\,.
\end{equation}

Again with the understanding that $\ED3 \to \ED3/c_2(\mathfrak{g}_D)$ for $D \in \{0,\ldots, 25\}$,
we can write
the full superpotential as
\begin{equation}\label{eq:wfullex}
    W= W_0 + \sum_{i=1}^{114} \text{exp}\biggl(-\int_{D_i}\ED3\biggr)\,,
\end{equation}
with $W_0$ given by \eqref{eq:w0value}.

Next, one can show that there
exists a point $t^{\star}$ in the K\"ahler cone of $K$ where
\begin{equation}
    \text{Vol}(D_1) = \cdots = \text{Vol}(D_{25}) = 6\,\text{Vol}(D_{26}) = \cdots = 6\, \text{Vol}(D_{114})\,,
\end{equation}
so that all 114 terms in Eq.~\eqref{eq:wfullex} are commensurate.  Moreover, at the given point $(z_\star^a,\tau_\star, t^{\star})$ in moduli space, all other contributions to the superpotential are parametrically suppressed compared to those listed, so that the vacuum structure is governed by \eqref{eq:wfullex}.

Finally, one can verify that the potential $\Veff$ for the K\"ahler moduli induced by the superpotential
Eq.~\eqref{eq:wfullex} has a stable  minimum
preserving $\mathcal{N}=1$ supersymmetry, and at which $\Veff<0$ \cite{Demirtas:2021nlu}.  Thus, $\Lambda<0$ and the solution is an AdS$_4$ vacuum.

\section{ Open questions }

This chapter has focused on some of the best-understood solutions of string theory with $D<10$: compactifications of superstring theory on Ricci-flat manifolds $\K$ with special holonomy.
But compactification of supergravity theories suggests the existence of a vast space of more general solutions, even restricting to supersymmetric ones.  For example, the $\IIbtheory$ solutions of \S\ref{s:ccy} are a special subclass
in which the effects of fluxes leave the metric on $\K$ conformally Ricci-flat, and on relaxing this restriction one finds $\mathcal{N}=1$ supersymmetric compactifications of $\IIbtheory$ in which $\K$ has $SU(3) \times SU(3)$ structure, but does not have $SU(3)$ holonomy \cite{Grana:2005sn}.

String theory also admits generalized compactifications in which $\K$ is not a manifold.  The best established and most studied case is to take
$\K$ to be a world-sheet superconformal field theory with appropriate central charge (the same as a sigma model with target $T^k$).
There are many other proposals, often using established concepts in string/M duality (S, T and U-duality) in novel
geometric ways, for example as monodromies of a fibration.

Hardly anything systematic is known about non-supersymmetric solutions of superstring theory.  Present understanding of the theory allows for the study of non-supersymmetric solutions in EFTs obtained from string compactifications, but not yet directly in string/M-theory.
Fundamental improvements in our knowledge of string theory would be a welcome aid,
but a less ambitious and still necessary direction involves advances in computing EFTs from string theory.  Application of string worldsheet methods, string field theory, and computational geometry can gradually reveal more of the structure of $\mathcal{N}=1$ supersymmetric compactifications, and  in time this may lead to an understanding of supersymmetry breaking.

In \S \ref{ss:superQ}
we mentioned the idea that QFT can be used to rederive and confirm certain compactification results.  As an example, suppose the EFT at scale $M$
contains an asymptotically free gauge theory subsector which is only weakly coupled to the rest; then one can use gauge theory
arguments to approximately derive its EFT at lower scales.  This is how QCD physics would emerge from a realistic string compactification.

As a much more far-reaching example, stable AdS$_d$ compactifications are believed to be dual to $(d-1)$-dimensional conformal field theories.
This is very well established in highly supersymmetric cases.  While there are no known examples of CFTs dual to AdS$_{4}$
compactifications with $\mathcal{N}=0$ or $\mathcal{N}=1$ supersymmetry (including those of \S \ref{s:typeII}),
there are no strong arguments against their existence and this is an outstanding question.
Another critical question is to find gauge duals or other alternative constructions of dS and Minkowski compactifications.

To conclude, compactifications of string theory are an extremely rich interface connecting geometry, quantum string/M-theory, and observable phenomena in particle physics and cosmology.
Progress over the past two decades has been continuous, with ongoing technical advances but with no revolutions in the mathematics or the physics, and much remains to be discovered.

\bibliographystyle{utphys}
\bibliography{references}
\end{document}